\documentclass[a4paper,11pt]{article}
\usepackage{jcappub} 
\usepackage{lineno}

\usepackage{graphicx}
\usepackage{dcolumn}
\usepackage{bm}
\usepackage{amsmath,amssymb} 
\usepackage{amsfonts}
\usepackage{subfigure}
\usepackage{graphicx}
\usepackage{ulem}
\usepackage{soul}
\usepackage{color}
\usepackage{xcolor}
\usepackage{float}
\usepackage{makecell}
\usepackage{newtxtext,newtxmath}
\usepackage{mathrsfs}
\usepackage{gensymb}
\usepackage{multirow}
\usepackage{booktabs}
\usepackage{threeparttable}
\usepackage{ulem}
\usepackage{rotating}
\usepackage{graphicx}
\usepackage{tabularx}
\usepackage{verbatim}
\usepackage{url}
\usepackage{hyperref}
\usepackage[utf8]{inputenc}

\title{The first AKRA mass map reconstruction from HSC Y1 data}

\UseRawInputEncoding

\author[a,c]{Yuan Shi}
\emailAdd{yshi@sjtu.edu.cn}

\author[a,b,c]{Pengjie Zhang}
\emailAdd{zhangpj@sjtu.edu.cn}
\author[a,c]{Zhao Chen}
\author[a,c]{Jian Qin}
\author[a,c]{Li Cui}
\author[d,e]{Furen Deng}
\author[g]{Ji Yao}

\affiliation[a]{Department of Astronomy, School of Physics and Astronomy, Shanghai Jiao Tong University, Shanghai, 200240, China}
\affiliation[b]{Division of Astronomy and Astrophysics, Tsung-Dao Lee Institute, Shanghai Jiao Tong University, Shanghai, 200240, China}
\affiliation[c]{Key Laboratory for Particle Astrophysics and Cosmology (MOE) / Shanghai Key Laboratory for Particle Physics and Cosmology, China}
\affiliation[d]{National Astronomical Observatories, Chinese Academy of Sciences, 20A Datun Road, Beijing 100101, P. R. China}
\affiliation[e]{University of Chinese Academy of Sciences, School of Astronomy and Space Science, Beijing 100049, P. R. China}
\affiliation[g]{Shanghai Astronomical Observatory (SHAO), Nandan Road 80, Shanghai, China}

\abstract{
Weak lensing mass-mapping from shear catalogs faces systematic challenges from survey masks and spatially varying noise. To overcome these issues and reconstruct unbiased convergence $\kappa$ maps, we have constructed the AKRA (Accurate Kappa Reconstruction Algorithm), a prior-free  and maximum-likelihood based analytical method. It has been validated  for mock shear catalogs with a variety of survey masks.
In this work, we present the first real-data application of the AKRA on the Subaru Hyper Suprime-Cam Year 1 (HSC Y1) data.
We first validate AKRA using mock shear catalogs from the \texttt{Kun} simulation suite, with masks corresponding to the six HSC Y1 regions (\texttt{GAMA09H}, \texttt{GAMA15H}, \texttt{HECTOMAP}, \texttt{VVDS}, \texttt{WIDE12H}, and \texttt{XMMLSS}). The investigated statistics, including the lensing power spectrum,  $\langle \kappa^2\rangle$, $\langle \kappa^3\rangle$, and the one-point probability distribution function of $\kappa$, are all unbiased.  
We then apply AKRA to the HSC Y1 shear catalog and provide reconstructed $\kappa$ maps ready for subsequent scientific analyses.
}

\begin{document}
\maketitle
\flushbottom

\section{Introduction} \label{sec:intro}
Weak gravitational lensing (WL) directly measures the projected mass distribution in the Universe, and is therefore a powerful probe of structure growth and cosmology \cite{Bartelmann2001, Refregier2003, Munshi2008,Fu2014, Kilbinger2015, Mandelbaum2018}.
Over the past decade, stage-III weak-lensing surveys such as the Dark Energy Survey (DES; \cite{Abbott2016, Amon2022_DES_Y3, Secco2022_DES_Y3}), the Hyper Suprime-Cam Subaru Strategic Program (HSC; \cite{Aihara2018, Hikage2019_HSCY1, Hamana2020_HSCY1}), and the Kilo-Degree Survey (KiDS; \cite{Kuijken2015, Asgari2021_KiDS1000}) have achieved S/N$\sim 30$ in cosmic shear measurements. 
The ongoing and upcoming stage-IV surveys will advance weak lensing into a new era, improving S/N  by an order of magnitude or more (e.g, \cite{Yao2024CSST}). These include Euclid \cite{Laureijs2011, Euclid2022}, the Vera C. Rubin Observatory Legacy Survey of Space and Time (LSST; \cite{LSST2009, Ivezi2020}), the Roman Space Telescope \cite{Spergel2015}, and the Chinese Space Station Telescope (CSST; \cite{Gong2019, Zhan2021, Yao2024CSST}).

So far, most weak-lensing analyses in galaxy surveys have been performed directly using  shear catalogs.
However, for a variety of applications it is more convenient to work with  convergence ($\kappa$) maps.
First, many non-Gaussian statistics are naturally defined on $\kappa$ fields.
These include one-point probability distribution functions (PDFs), peak and void counts \citep{Shan2018,Martinet2018,Liu2023}, skewness and kurtosis \citep{VanWaerbeke2013,Petri2015,Chang2018}, Minkowski functionals \citep{Kratochvil2012,Vicinanza2019,Zurcher2021} and  scattering transform \citep{Cheng2020,Cheng2021}. Furthermore, both the conventional three-point correlation function and bispectrum \citep{Takada2003,Fu2014} and the recent techniques such as  machine-learning \citep{Gupta2018,Ribli2019a,Fluri2022,LuTianhuan2023}, and field-level inference \citep{Zhou2024,Zeghal2025} are more convenient to implement using the $\kappa$ maps instead of shear catalogs. 
Second, in galaxy–galaxy lensing, the tangential shear $\gamma_t(\theta)$ is inherently nonlinear, as it includes contributions from scales $\theta'< \theta$.
This intrinsic nonlinearity complicates the theoretical modeling \citep{Baldauf2010,MacCrann2020,Park2021,Prat2022}. In contrast, the $\kappa$-galaxy cross-correlation is free of this issue. 

To convert a shear catalog into a convergence map, the Kaiser-Squires (KS) method \cite{Kaiser1993, VanWaerbeke2013, Vikram2015, Gatti2022} is widely used. However, it suffers from biases in the presence of survey masks and spatially varying noise.
Various alternative approaches to the KS method have been developed, including Bayesian forward modeling \cite{Alsing2016, Alsing2017, Porqueres2022}, Wiener filtering \cite{Jeffrey2018}, sparsity- and lognormal-inspired methods \cite{Leonard2014, Price2019, Jeffrey2018, Fiedorowicz2022b, Fiedorowicz2022a}, and wavelet reconstructions \cite{Starck2006, Starck2021}. 
These methods typically impose explicit assumptions about the statistical properties of the convergence field.
Alternatively,  we build the Accurate Kappa Reconstruction Algorithm (AKRA,   \citep{Shi2024,Shi2025}). It is a prior-free and maximum-likelihood based  analytical method, applicable to both flat and curved sky. Tests against various mock shear catalogs show that it delivers stable and unbiased $\kappa$ reconstructions .

In this work, we present the fist application of AKRA to real data, using the first-year shear catalog from the Hyper Suprime-Cam Subaru Strategic Program (HSC Y1; \cite{Mandelbaum2018_HSCY1_data}), which covers 136.9 deg$^2$ over six disjoint fields. 
This catalog has been widely used in recent studies of non-Gaussian weak-lensing statistics, including peak counts, Minkowski functionals, PDFs, and scattering transforms \cite{Liu2023_HSC, Lu2023_HSC, Thiele2023_HSCY1_PDF, Grandón2024_HSC, Marques2024_HSC, Novaes2024_HSC, Armijo2025_HSC, Cheng2025_HSC}. All of these studies have so far relied on the KS reconstructed convergence maps. Therefore $\kappa$ maps reconstructed using methods other than KS would serve as independent checks for the above statistical analysis. AKRA passes all investigated tests against mock shear catalogs with HSC Y1 survey boundary, masks and spatially varying noise.  Therefore we believe that the AKRA reconstructed mass maps are ready for subsequent scientific analysis. All maps in this work will be made publicly available upon the publication of this paper.

The paper is organized as follows. In Section~\ref{sec:method}, we present the formulation of AKRA and describe the mass map reconstruction pipeline.
Section~\ref{sec:HSC_Y1} introduces the HSC Y1 data set and presents the reconstructed convergence maps. 
In Section~\ref{sec:simulation}, we describe the simulation framework and validation tests.
Section~\ref{sec:discussion} provides our discussions and conclusions.

\section{Method} \label{sec:method}

\begin{figure}[h!]
    \centering
    \includegraphics[width=0.6\columnwidth]{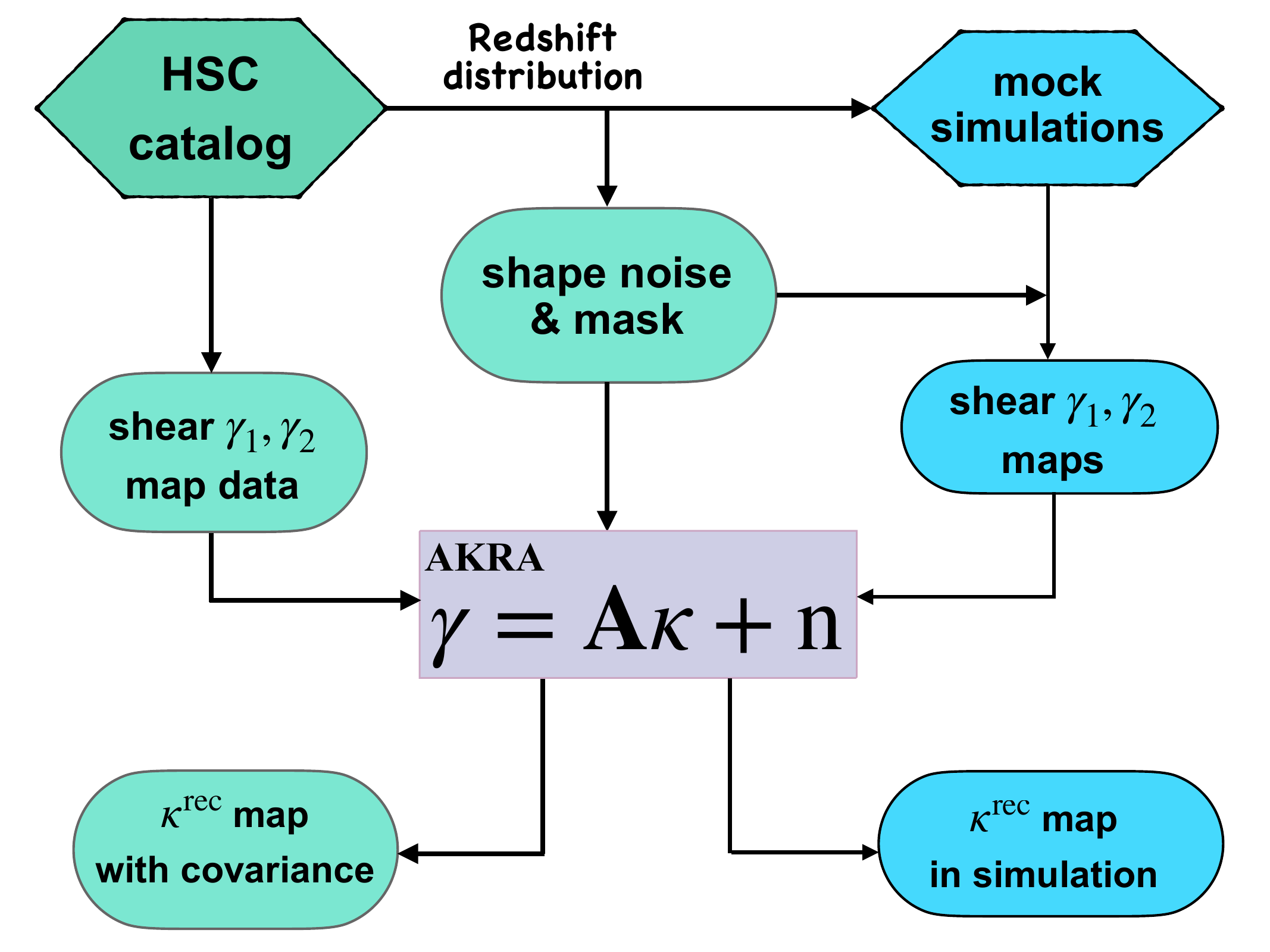}
    \caption{Overview of the mass map reconstruction pipeline. The left side shows the processing of HSC Y1 galaxy catalogs into shear maps and masked fields, while the right side shows parallel validation using mock simulations. The central AKRA module performs prior-free reconstruction of the convergence field.}
    \label{fig:pipeline}
\end{figure}

In this work, we structure the analysis into two parts: (i) an observational part that reconstructs $\kappa$ from the HSC~Y1 shear catalog(Sec.~\ref{sec:method}); and (ii) a simulation part that creates HSC-like mocks to validate the reconstruction and quantify uncertainties (Sec.~\ref{sec:simulation}). Figure~\ref{fig:pipeline} summarizes the workflow.

The shear-convergence relation in the presence of masks and noise can be written as a linear system:
\begin{equation}
\boldsymbol{\gamma} = \mathbf{A}\,\boldsymbol{\kappa} + \boldsymbol{n}\ .
\end{equation}
For HSC we can adopt the flat sky approximation and use the flat sky version of AKRA \cite{Shi2024}.  Then $\boldsymbol{\gamma} = \begin{bmatrix} \tilde{\gamma}_{1}^m(\vec{L}) \\ \tilde{\gamma}_{2}^m(\vec{L}) \end{bmatrix}$
is the observed (masked) shear, \(\boldsymbol{\kappa}\) is the true convergence, and \(\boldsymbol{n}\) is noise. The linear operator \(\mathbf{A}\) 
The operator \(\mathbf{A}\) combines (i) the geometric lensing kernels, which project \(\kappa\) into \((\gamma_1,\gamma_2)\) with angular factors \(\cos 2\phi_\ell\) and \(\sin 2\phi_\ell\) for each Fourier mode \(\boldsymbol{\ell}\), and (ii) the mask-induced mode coupling through a Fourier-space mask operator \(\mathbf{M}\).\footnote{In the idealized case without masks (and noise-free), \(\mathbf{M}\to\mathbf{I}\),  AKRA reduces to KS ~\cite{Shi2024,Shi2025}.}
$$\mathbf{A} = \begin{bmatrix}
\cos \left(2 \phi_{\ell_{1}}\right) \mathbf{M} \\
\sin \left(2 \phi_{\ell_{1}}\right) \mathbf{M}\ 
\end{bmatrix}.$$
The associated covariance of the estimator is then:
\begin{equation}
\mathbf{C} = \left( \mathbf{A}^{\rm T} \mathbf{N}^{-1} \mathbf{A} \right)^{-1},
\end{equation}
where $\mathbf{N} \equiv \langle \mathbf{n}\,\mathbf{n}^{\mathbf T} \rangle$ is the noise covariance matrix.  $\mathbf{C}$ is often ill-conditioned, especially in the presence of large mask regions or incomplete sampling of Fourier modes. To regularize the inversion and suppress noise amplification, we adopt a small regularization matrix \(\mathbf{R} = \lambda \mathbf{I}\) where $\lambda \sim 10^{-3}$. The estimator then becomes:
\begin{equation}
\hat{\boldsymbol{\kappa}} = \left( \mathbf{A}^{\rm T} \mathbf{N}^{-1} \mathbf{A} + \mathbf{R} \right)^{-1} \mathbf{A}^{\rm T} \mathbf{N}^{-1} \boldsymbol{\gamma}.
\label{eq:kappa_hat}
\end{equation}

In our previous work \cite{Shi2024,Shi2025}, a simplified model with $\mathbf{N}^{-1} = \mathbf{I}$  was assumed, effectively treating the noise as homogeneous and uncorrelated. In this study, we go beyond this approximation by explicitly constructing $\mathbf{N}$ based on the spatially varying shear noise across the field, derived from the weighted shape noise  in each pixel.
\begin{figure*}[h!]
    \centering
    \includegraphics[width=0.65\textwidth]{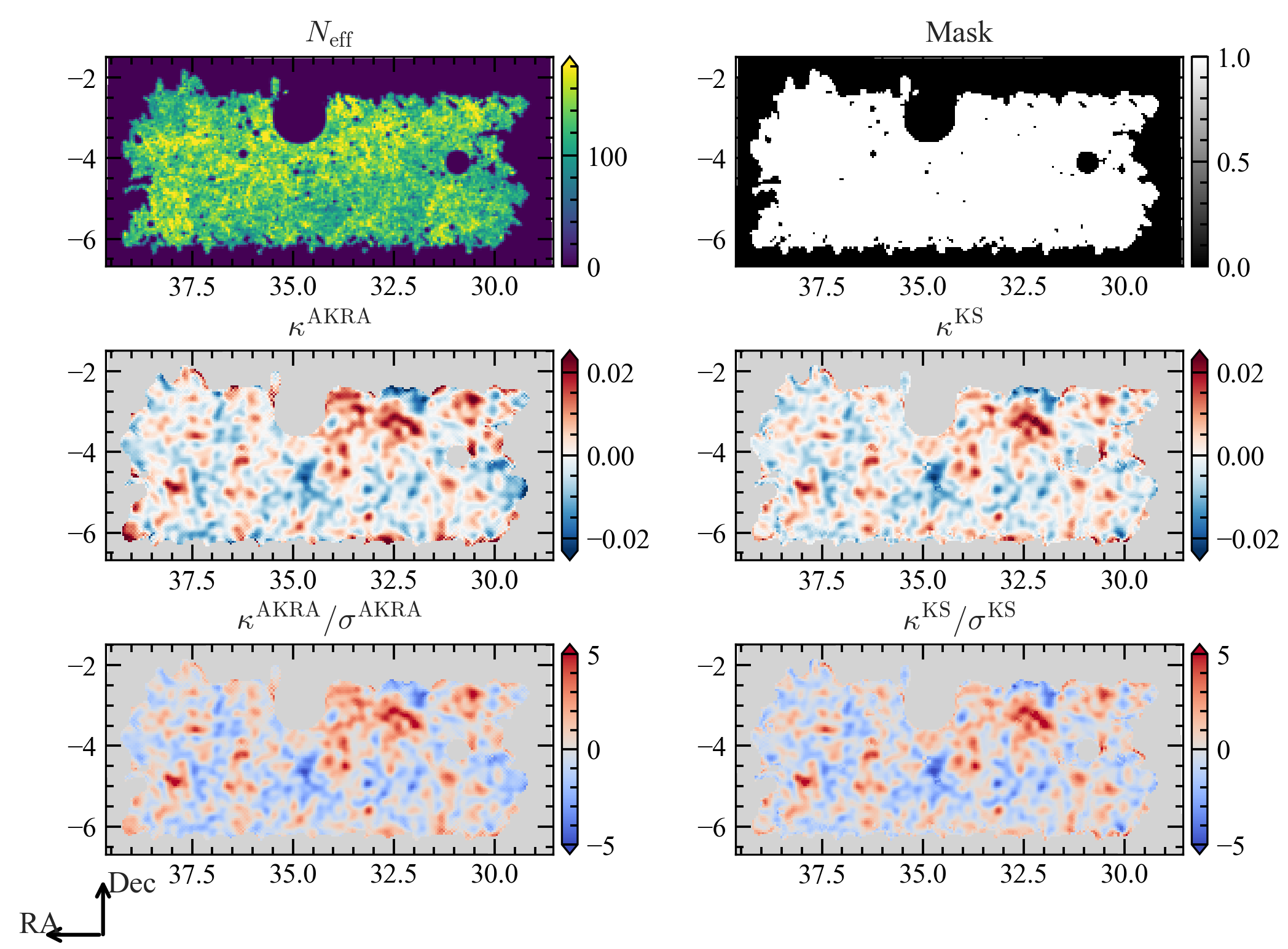}
    \caption{
    Weak lensing mass mapping results for the \texttt{XMMLSS} field with source galaxies in $z_{\text{best}}\in(0.3,1.5]$. 
    Panels show (from left to right, top to bottom): 
    (1) the effective source number density $n_\mathrm{eff}$; 
    (2) the binary mask derived from $n_\mathrm{eff}$ (pixels with $n_\mathrm{eff}<10$ galaxies are excluded); 
    (3) the AKRA reconstructed convergence $\kappa^{\rm AKRA}$; 
    (4) the KS convergence $\kappa^{\rm KS}$; 
    (5) the SNR map of $\kappa^{\rm AKRA}$; 
    (6) the SNR map of $\kappa^{\rm KS}$. 
    The convergence, and SNR maps are smoothed with a Gaussian window of $\sigma = 5$ arcmin for better demonstration.
    }
    \label{fig:kappa_XMMLSS}
\end{figure*}

\begin{figure*}[!htbp]
    \centering
    \includegraphics[width=0.99\textwidth]{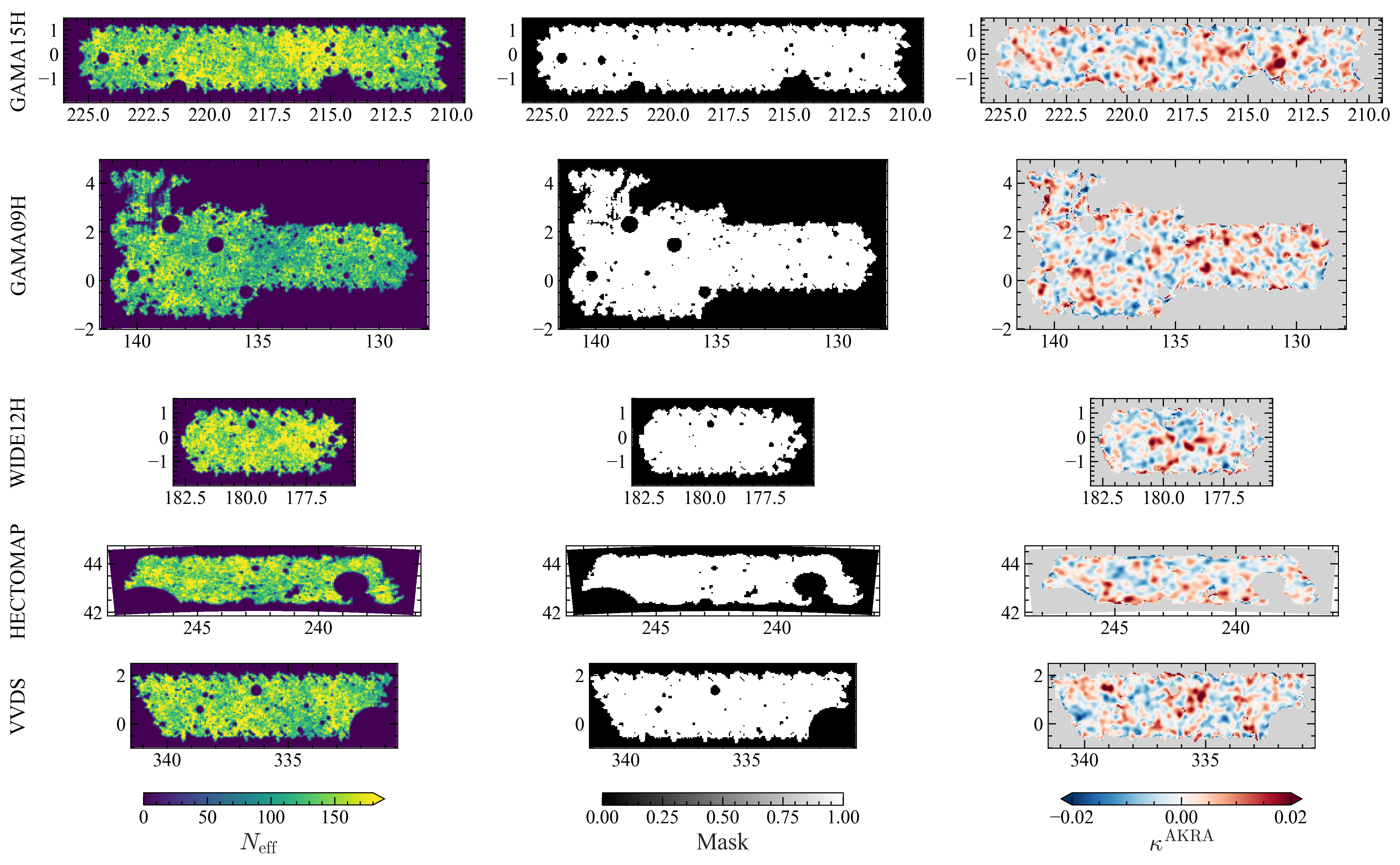}
    \caption{
    Overview of mass mapping results in the five representative HSC-Y1 fields: 
    \texttt{GAMA15H}, \texttt{GAMA09H}, \texttt{WIDE12H}, \texttt{HECTOMAP}, and \texttt{VVDS}. 
    For each field, panels (from left to right) show the effective source density $n_\mathrm{eff}$, the binary mask, and the AKRA reconstructed convergence maps $\kappa^{\rm AKRA}$. 
    The convergence maps are smoothed with a Gaussian window of $\sigma = 5$ arcmin for demonstration. 
    }
    \label{fig:hsc_kappa_map}
\end{figure*}

\section{HSC Y1 Data Results} \label{sec:HSC_Y1}

The first-year data release of the HSC~Y1 shear catalog \cite{Mandelbaum2018_HSCY1_data} covers an area of 136.9~deg$^2$ across six disjoint Wide fields: \texttt{GAMA09H}, \texttt{GAMA15H}, \texttt{HECTOMAP}, \texttt{VVDS}, \texttt{WIDE12H}, and \texttt{XMMLSS}. It contains shape measurements and shear estimates for approximately 12.1~million galaxies. We adopt
Photometric redshifts (photo-z) from the best estimate of \texttt{Ephor~AB} algorithm \citep{Tanaka2018}, denoted as $z_{\mathrm{best}}$.
Sources are restricted to the range $0.3 < z_{\mathrm{best}}\le 1.5$.
\footnote{The public HSC~Y1 catalog also provides six alternative photo-$z$ estimates per galaxy: \texttt{MLZ, Ephor, Mizuki, NNPZ, Frankenz, DEmP}. We choose \texttt{Ephor~AB} following the HSC~Y1 cosmology analyses \citep{hikage_cosmology_2019,hamana_cosmological_2020}. } The effective number of source galaxies in a given pixel is defined as
$$
N_{\mathrm{eff}} = \sum_{i \in \mathrm{pix}}
\frac{\sigma_{\mathrm{SN},i}^2}{\sigma_{\mathrm{SN},i}^2 + \sigma_{m,i}^2}\ ,
$$
where $\sigma_{\mathrm{SN},i}$ and $\sigma_{m,i}$ denote the intrinsic shape noise and measurement error of galaxy $i$, respectively. The average effective source density $n_{\mathrm{eff}} = N_{\mathrm{eff}}/\Omega  \simeq 17~\mathrm{arcmin}^{-2}$. The corresponding shear noise variance is given by $\sigma_{\mathrm{pix}}^2 =\sigma_\gamma^2/N_{\mathrm{eff}}$, with $\sigma_\gamma \simeq 0.24$ being the intrinsic shape noise of HSC~Y1 source galaxies.


For the reconstruction, we adopt a flat-sky grid with 3~arcmin pixels. Fig. \ref{fig:kappa_XMMLSS} shows the $N_{\rm eff}$ map of the \texttt{XMMLSS} region, which is highly inhomogeneous. Together with the complicated survey boundary, they represent a major issue to be dealt with. We also need to define the binary mask $m(\vec{\theta})$. $m=1$ if $N_{\rm eff}>N_{\rm thres}$ and $0$ otherwise. AKRA naturally handles inhomogeneous $N_{\rm eff}$ distributions through its noise weighting scheme; the threshold is only required to exclude pixels with extremely low galaxy counts where the reconstruction becomes unreliable. For a pixel size of $3'\times3'$, the average $N_{\rm eff}$ is $\bar{N}_{\rm eff} = 153$. Most mass mapping studies adopted a threshold of $0.5 \times \bar{N}_{\rm eff}$, which would correspond to $N_{\rm thres} \sim 75$ for our case, discarding $\sim 6\%$ of galaxies. We adopt $N_{\rm thres}=10$, which is much smaller than $\bar{N}_{\rm eff}/2$. This conservative cut discards only $\sim0.1\%$ of galaxies (i.e., $\sim 10^4$ out of the total $\sim 12$ million galaxies)\footnote{For reference, more aggressive thresholds would discard substantially more information (1.4\% for $N_{\rm thres}=50$, 6.3\% for $N_{\rm thres}=100$, and $\sim$50\% for $N_{\rm thres}=150$).}, while ensuring numerical stability in the reconstruction. Fig. \ref{fig:kappa_XMMLSS} shows the mask map in the \texttt{XMMLSS} region.

For the reconstruction we adopt a small diagonal regularization term\footnote{In the idealized case without masks, $\mathbf{M} \to \mathbf{I}$, the largest eigenvalue of $\mathbf{A}^{\rm T} \mathbf{A}$ is unity since $\mathbf{A}^{\rm T} \mathbf{A} = \text{diag}(\cos^2 2\phi_{\ell}) + \text{diag}(\sin^2 2\phi_{\ell}) = \mathbf{I}$. In the presence of masks, some eigenvalues become small due to mode coupling, leading to numerical instability in the inversion. Our choice of $\lambda = 10^{-3}$ relative to the largest eigenvalue keeps the inversion stable while maintaining the systematic error below 0.1\% (see Ref.~\cite{Shi2024} for a detailed eigenvalue analysis).}$\mathbf{R}=\lambda\mathbf{I}$ with $\lambda=10^{-3}$.
Noise properties  are characterized using 300 randomized shear realizations obtained by randomizing galaxy orientations while keeping their positions and weights fixed. 
The pixel-level noise of the reconstructed convergence field, $\sigma_\kappa$, is then estimated from the variance of reconstructed $\kappa$ across these realizations.

\begin{figure*}[h!]
    \centering
    \includegraphics[width=0.85\textwidth]{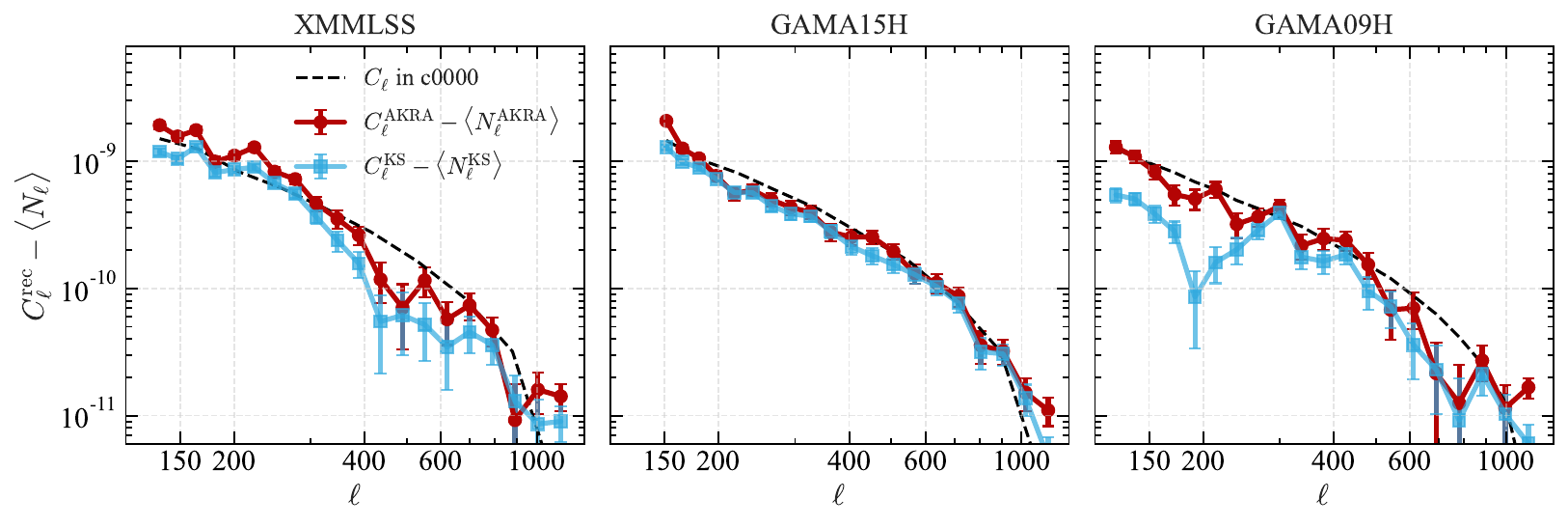}
    \caption{
        Power spectrum of the reconstructed convergence maps from HSC Y1 data in the redshift range $0.3<z_s\leq1.5$. From left to right, we show results for three large fields: \texttt{XMMLSS}, \texttt{GAMA15H} and \texttt{GAMA09H}.
        The error bars are estimated from 300 noise realizations per field.
        The black dashed curves indicate the theoretical power spectrum for a fiducial $\Lambda$CDM cosmology, shown for reference rather than as a best-fit model.
    }
    \label{fig:data_power_spectrum}
\end{figure*}

\subsection{Mass mapping results}
We then apply AKRA to all six HSC~Y1 fields, 
yielding the first set of convergence ($\kappa$) maps from this method on real survey data. 
Figs.~\ref{fig:kappa_XMMLSS} and \ref{fig:hsc_kappa_map} illustrate the complex survey masks and strong spatial variations in source density across the HSC~Y1 fields, together with the reconstructed $\kappa$ maps.

As a representative example, we highlight the \texttt{XMMLSS} region in Fig.~\ref{fig:kappa_XMMLSS}.
For comparison, we also perform KS reconstructions on the same grids. 
In well-sampled interior regions, AKRA and KS recover consistent large-scale structures, and the overall pattern of peaks and voids agrees. 
Near survey boundary and around masks, the reconstructions differ more significantly. 
Since the shear-convergence relation is non-local in real space, masks in the shear catalog impact the reconstruction of convergence in the unmasked regions in KS method.  
In contrast, AKRA incorporates the survey mask directly into a prior-free maximum-likelihood framework, yielding minimum-variance estimates of the convergence field under realistic survey conditions.
The accuracy of AKRA in handling mask effects has been validated with dedicated simulations in previous work \citep{Shi2024}.


The corresponding power spectra are presented in Fig.~\ref{fig:data_power_spectrum}; error bars are estimated from 300 noise realizations per field. 
These $\kappa$ maps and their power spectra constitute the primary data products released in this work and provide the basis for subsequent scientific analyses.
Here we focus on presenting the reconstructed maps and derived data products from real HSC~Y1 observations. 
A full cosmological interpretation from these results will require rigorous treatment of observational systematics, including photo-z uncertainties, shear calibration, and intrinsic alignments, to be addressed in future work.

\section{Validation with Simulations} \label{sec:simulation}


To quantify the accuracy of AKRA reconstruction of HSC~Y1, we perform validation tests using  mock simulations with known input convergence fields. They allow us to evaluate both the statistical accuracy of the reconstructed maps and examine whether AKRA faithfully reproduces the non-Gaussian characteristics of the underlying matter field.

The mock convergence fields are generated from the high-resolution \textsc{kun} simulation suite~\citep{2025SCPMA..6889512C}, which is part of the \textsc{Jiutian} program designed in preparation for the forthcoming CSST cosmological survey~\citep{2025SCPMA..6809511H}. 
The \textsc{kun} suite consists of 129 $N$-body simulations spanning an eight-dimensional cosmological parameter space, including models with dynamic dark energy and massive neutrinos. 
It has been widely used to build cosmological emulators for the matter power spectrum, halo mass function, and biased tracer spectra~\citep{2025SCPMA..6889512C,2025SCPMA..6809513C,2025arXiv250604671Z}. 
In this work, we adopt the fiducial realization (c0000) from the suite, with cosmological parameters 
$\Omega_M=\Omega_\mathrm{cb}+\Omega_{\nu}=0.3111$, 
$\Omega_b=0.0490$, 
$H_0=67.66~\mathrm{km~s^{-1}~Mpc^{-1}}$, 
$n_s=0.9665$, 
$\sigma_8=0.81$, 
and a dark energy equation of state of $w=-1$, $w_a=0$, 
with a total neutrino mass of $\Sigma m_\nu = 0.06~\mathrm{eV}$.
Convergence maps are generated by stacking the full-sky projected density maps with thickness $50\,h^{-1}\mathrm{Mpc}$ in the on-the-fly lightcone.
The corresponding shear components $(\gamma_1, \gamma_2)$ are then obtained.
The mock simulations also adopt the same photo-$z$ distribution as the HSC~Y1 data.
We emphasize that throughout this section we adopt only the fiducial cosmology of the \textsc{kun} suite, without exploring the full parameter space.

For each of the six HSC~Y1 fields, we generate 200 realizations that reproduce the survey geometry, pixelization, binary mask, and spatially varying effective number map $N_{\rm eff}(\vec\theta)$ measured from the data (Section~\ref{sec:HSC_Y1}). 
Each realization includes the input cosmological shear signal $(\gamma_1, \gamma_2)$, to which we apply the survey mask and add pixel-dependent shape noise.
We note that the observed ellipticity corresponds to reduced shear $g = \gamma/(1-\kappa)$ rather than shear $\gamma$ (see Eqs. (24) and (25) in Ref.~\cite{Shirasaki2019} for the exact relation). In the weak lensing regime where $\kappa \ll 1$, this distinction has a negligible impact on our results.
We reconstruct the convergence maps using both the AKRA and KS methods with the same settings as applied to the data.
Noise-only maps (without cosmological signal) are also generated in the same way to estimate noise biases.

\begin{figure*}[h!]
    \centering
    
    \includegraphics[width=0.88\textwidth]{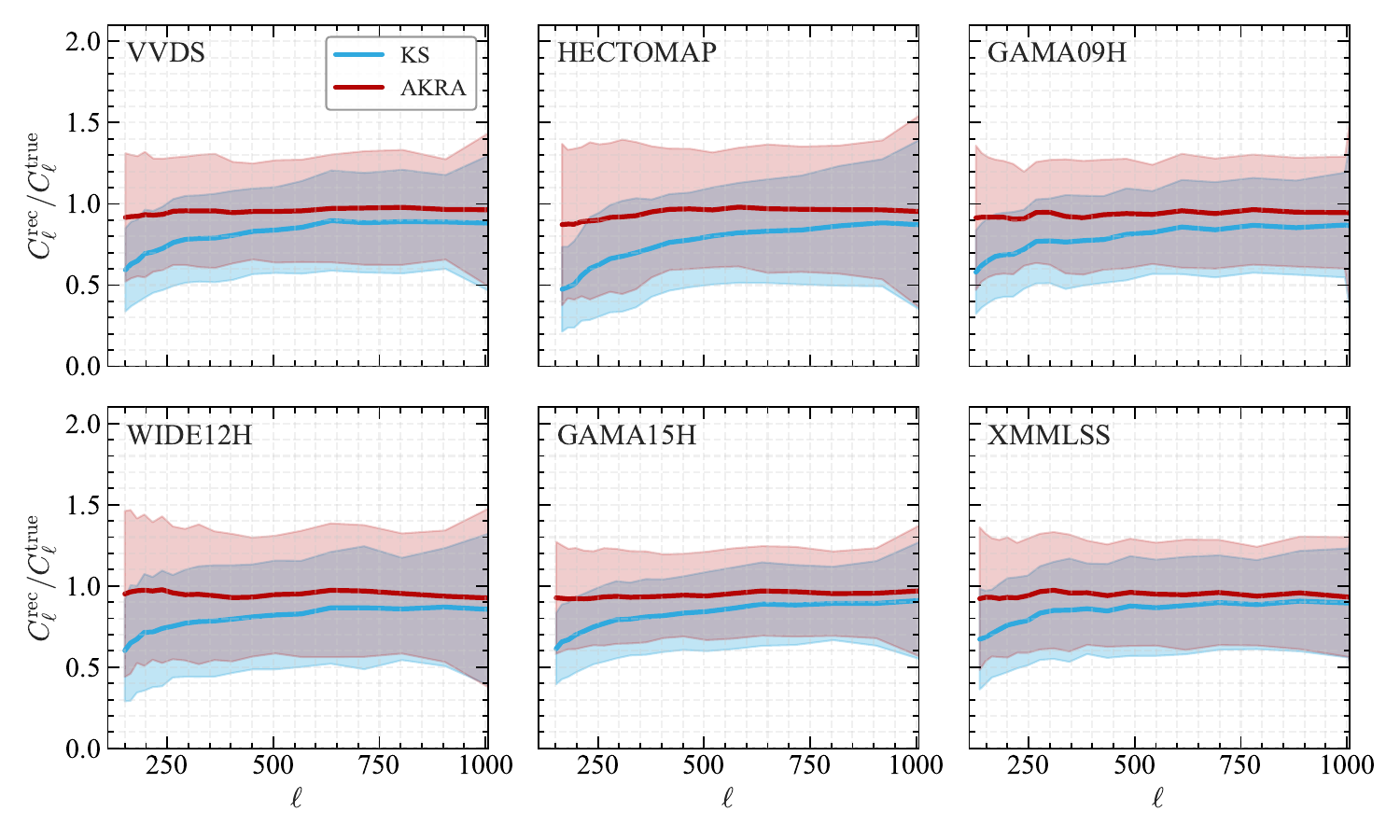}
    \makebox[\textwidth][c]{\textbf{(a)}}
    \vskip 0.3cm
    \includegraphics[width=0.88\textwidth]{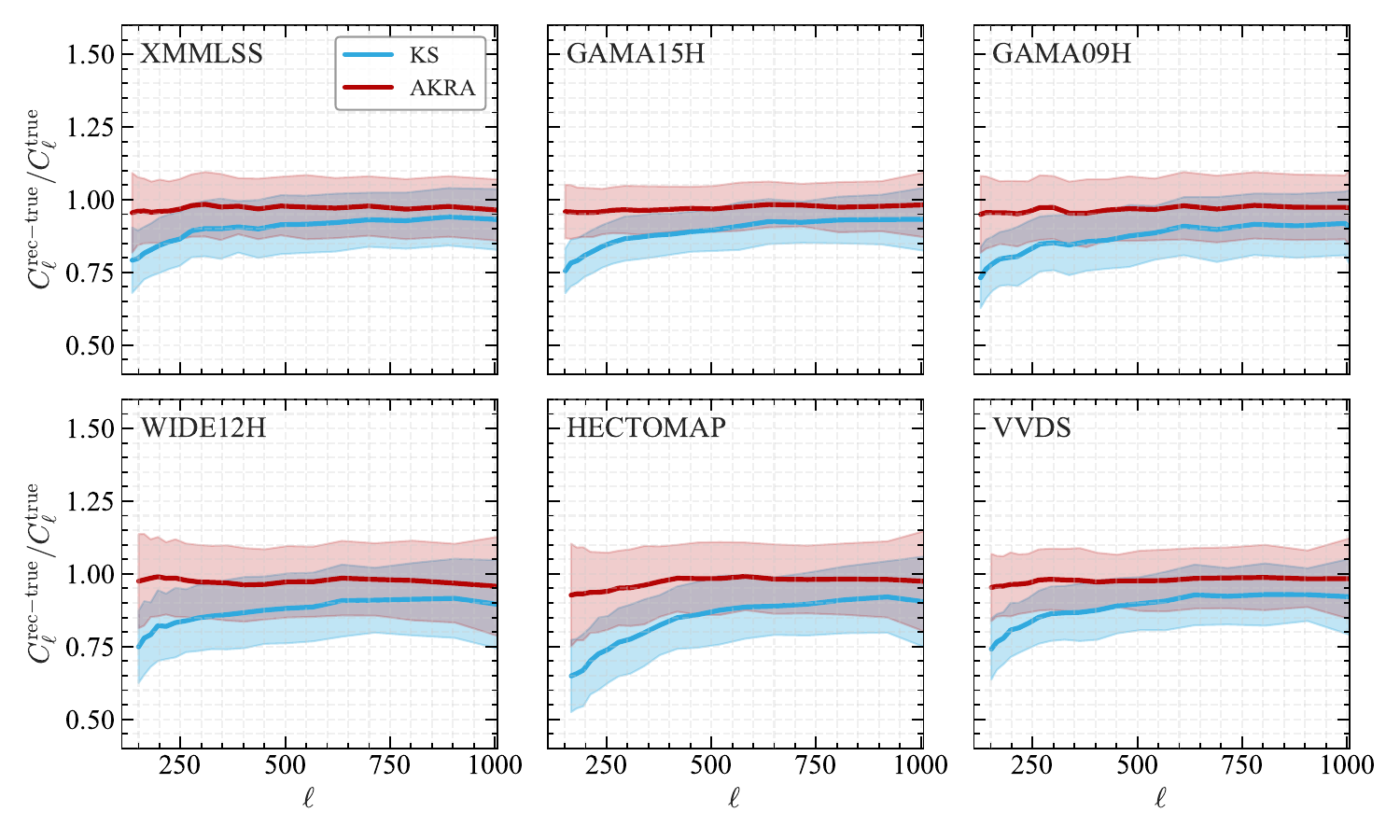}
    \makebox[\textwidth][c]{\textbf{(b)}}

    \caption{
        Validation of reconstructed convergence power spectra from mock simulations. 
        Panel (a) shows the ratio of noise-subtracted auto-spectra $C_\ell^{\rm rec}/C_\ell^{\rm true}$, 
        and panel (b) shows the ratio of cross-spectra $C_\ell^{\rm rec\!-\!true}/C_\ell^{\rm true}$ between reconstructed and true fields. 
        Results are averaged over 200 realizations per field, with error bars indicating the standard deviation. 
        AKRA (in red) follows the true input spectrum across a broad multipole range, 
        demonstrating that its treatment of masks and spatially varying noise preserves both large- and small-scale modes. 
        KS (in blue), in contrast, underestimates power on all scales, especially on large scales.
        This difference illustrates that explicitly modeling the mask, as done in AKRA, is essential for an unbiased recovery of the convergence field.  
    }
    \label{fig:power_spectrum}
\end{figure*}


\subsection{Power spectrum results}
The angular power spectrum provides a direct two-point diagnostic of reconstruction accuracy.
For each mock, we measure (i) the auto-spectrum of the reconstructed convergence map and (ii) its cross-spectrum with the input (“true”) convergence over the unmasked region.
We quantify performance using the ratios
\[
\frac{C_\ell^{\rm rec}}{C_\ell^{\rm true}}, \qquad 
\frac{C_\ell^{\rm rec\!-\!true}}{C_\ell^{\rm true}},
\]
where $C_\ell^{\rm true}$ is the spectrum of the true field, $C_\ell^{\rm rec}$ is the noise-subtracted auto-spectrum of the reconstruction, and $C_\ell^{\rm rec\!-\!true}$ is the cross-spectrum.
Amplitude biases in $C_\ell^{\rm rec}$ can, in principle, be corrected by rescaling individual $\ell$-modes, whereas phase errors probed by $C_\ell^{\rm rec\!-\!true}$ cannot.
This makes achieving an unbiased $C_\ell^{\rm rec\!-\!true}/C_\ell^{\rm true}$ the more challenging requirement.

As shown in Fig.~\ref{fig:power_spectrum}, AKRA shows unbiased recovery over $200 \lesssim \ell \lesssim 1000$: both ratios are consistent with unity within $ 5\%$ when averaged over realizations, demonstrating accurate reconstruction of both amplitude and phase information. 
For reference, the total signal-to-noise ratio (SNR) of HSC~Y1 cosmic shear 
measurements is approximately $16$--$19$ \citep{Hikage2019_HSCY1, Hamana2020_HSCY1}, 
corresponding to a statistical uncertainty of $\sim 5$--$6\%$. 
Future surveys such as HSC~Y3 (SNR~$\sim 26$) and Stage~IV (SNR~$\gtrsim 100$) surveys will achieve significantly higher precision.
In contrast, KS exhibits a systematic $\sim 20\%$ suppression of power across all scales, with the largest deficit on large scales where mask-induced mode mixing is strongest.
In previous tests \citep{Shi2024,Shi2025}, similar analyses were performed using idealized mocks that included survey masks but neglected shape noise. 
Here we extend the validation to realistic conditions by incorporating both irregular masks and spatially varying noise, 
which together constitute the dominant observational challenges for weak-lensing mass mapping.

%

Crucially, the unbiased recovery demonstrated here will become even more valuable in future wide-field surveys (e.g., LSST, Euclid, CSST).
With higher source densities and larger sky coverage, these surveys will achieve higher S/N and make AKRA a promising tool for producing reliable convergence maps.

\begin{figure*}[h!]
    \centering
    \includegraphics[width=0.8\textwidth]{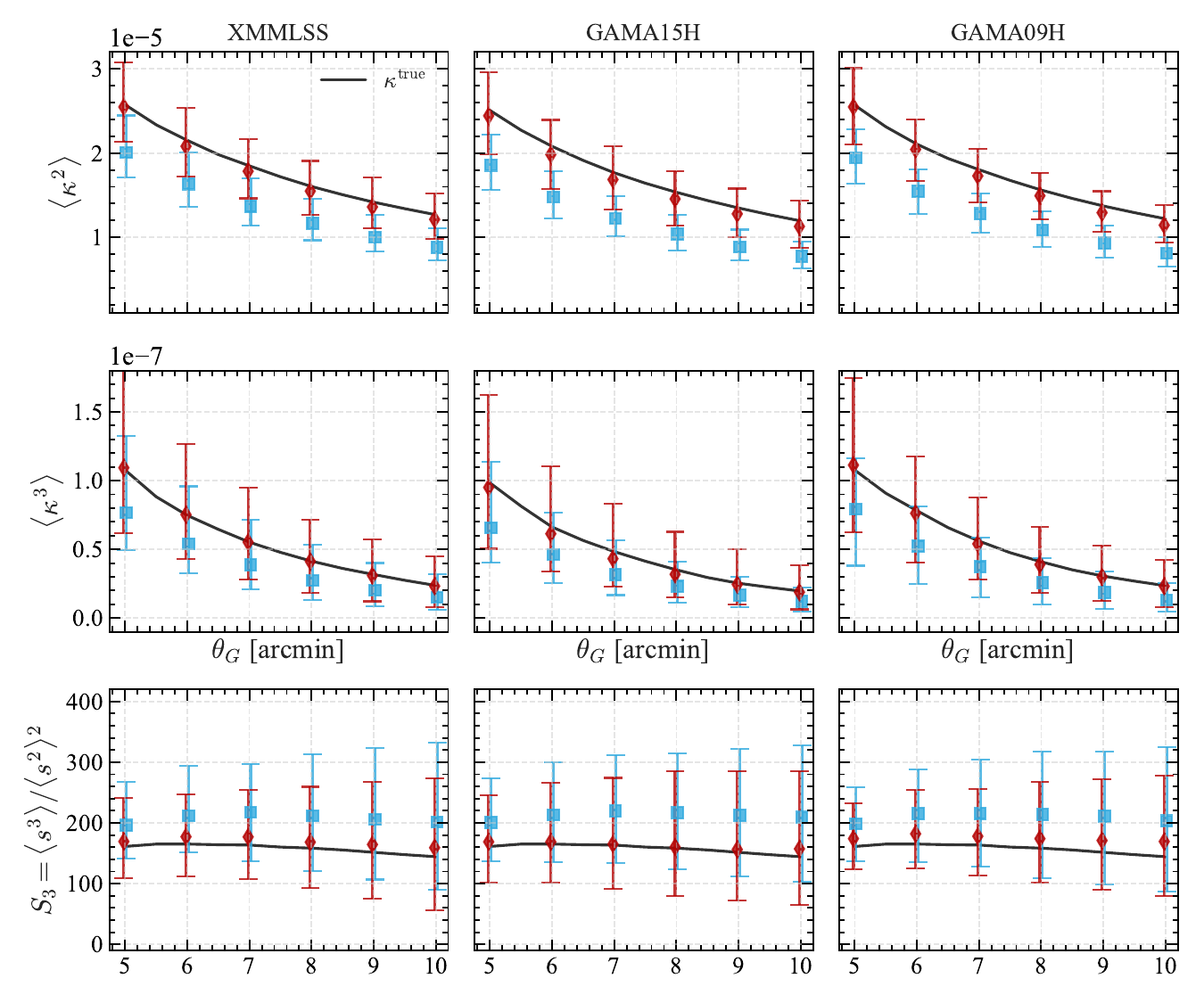}
    \caption{
    Higher-order statistics of reconstructed convergence fields from mock simulations in three representative HSC~Y1 regions (\texttt{XMMLSS}, \texttt{GAMA15H}, \texttt{GAMA09H}). 
    From top to bottom: the second moment $\langle\kappa^2\rangle$, the third moment $\langle\kappa^3\rangle$, and the normalized skewness $S_3=\langle\kappa^3\rangle/\langle\kappa^2\rangle^2$. 
    The x-axis shows the Gaussian smoothing scale $\theta_G \in \{5,6,7,8,9,10\}$~arcmin. 
    Black curves are measured from the true input maps; blue squares and red diamonds correspond to KS and AKRA reconstructions, respectively. 
    Error bars indicate the scatter across 200 mock realizations per field. 
    AKRA closely reproduces the true higher-order moments for every $\theta_G$ considered, in contrast to KS, which systematically biased the signal at all scales.
    These tests demonstrate that AKRA preserves non-Gaussian features of the convergence field, enabling analyses beyond two-point statistics.
    }
\label{fig:skewness_results}
\end{figure*}

\begin{figure*}[h!]
    \centering
    \includegraphics[width=0.8\textwidth]{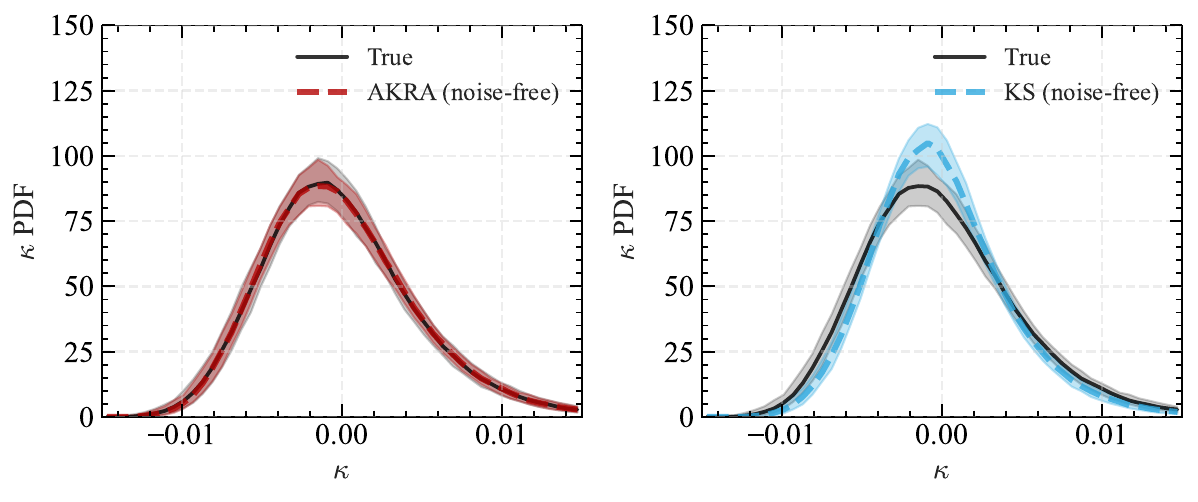}
    \caption{PDFs of reconstructed convergence fields from mock simulations in the representative HSC~Y1 region \texttt{XMMLSS}. Here we show noise-free reconstructions with the real survey mask applied (same pixelization and mask as in the data.)
    As we can see from the figure, AKRA closely matches the true PDF and preserves the non-Gaussian features of the field. In contrast, the KS reconstruction is biased: the PDF is enhanced near $\kappa\!\approx\!0$ and the tails are suppressed.
    Due to the limited S/N of HSC~Y1, we do not show noisy PDFs here; a fair comparison would require additional noise deconvolution.}
    \label{fig:pdf_results}
\end{figure*}

\subsection{Skewness and PDF results}
The mass maps become increasingly non-Gaussian on small scales due to non-linear structure growth. 
Higher-order statistics are therefore needed to capture this information. 
In this work, we use the skewness and one-point probability distribution function (PDF) as field-level diagnostics of whether AKRA preserves the non-linear features of the convergence field.

\subsubsection{Skewness}
Skewness, defined as the normalized third moment \cite{Bernardeau1997},
\begin{equation}
    S_3 = \frac{\langle \kappa^3 \rangle}{\langle \kappa^2 \rangle^{2}},
\end{equation}
and quantifies the asymmetry of the convergence distribution.
In perturbation theory, $S_3$ arises from non-linear mode coupling driven by gravitational instability and depends only weakly on the overall fluctuation amplitude, while receiving significant contributions from the high-$\kappa$ tail associated with rare massive halos.


We compute the second and third moments from AKRA reconstructions in three representative HSC~Y1 regions, estimating and subtracting the noise bias using 200 noise-only realizations.
Figure~\ref{fig:skewness_results} shows that AKRA accurately recovers the full dynamic range of both $\langle \kappa^2 \rangle$ and $\langle \kappa^3 \rangle$ across all smoothing scales considered, and hence yields unbiased estimates of $S_3$.
In contrast, KS systematically underestimates both moments and consequently overestimates $S_3=\langle\kappa^3\rangle/\langle\kappa^2\rangle^2$, typically requiring additional calibration to correct its amplitude bias.
The excellent agreement between AKRA and the truth demonstrates that AKRA preserves the non-Gaussian information encoded in skewness.

\subsubsection{PDFs}

The one-point convergence PDF, $P(\kappa)$, provides a complementary characterization of non-Gaussianity. 
Its high-$\kappa$ tail is dominated by the one-halo contribution from massive halos, while its negative tail arises from large-scale underdensities and voids. 
This makes $P(\kappa)$ sensitive to both the amplitude of matter fluctuations and the growth of non-linear structure.  
Theoretical descriptions include large-deviation theory \citep{barthelemy_probability_2021,boyle_nuw_2021,barthelemy_making_2024} and phenomenological lognormal models \citep{taruya_lognormal_2002}.

We measure the PDF from reconstructed $\kappa$ maps in noise-free mocks to isolate reconstruction effects. 
Figure~\ref{fig:pdf_results} shows the convergence PDFs in the representative \texttt{XMMLSS} field. We compare the PDFs of the true, AKRA-reconstructed, and KS-reconstructed $\kappa$ maps under noise-free conditions, applying the real survey mask.
The KS reconstruction shows a clear bias: the PDF is enhanced near $\kappa\approx 0$, while both the positive and negative tails are suppressed. In contrast, AKRA closely reproduces the shape of the true PDF, including the extended positive and negative tails characteristic of the non-Gaussian convergence field, as shown in Fig.~\ref{fig:pdf_results}.

%

In real data, intrinsic galaxy ellipticities introduce shape noise that propagates into the convergence field.
Including intrinsic shape noise would broaden the PDF and suppress its non-Gaussian features, making it difficult to clearly evaluate the differences between the AKRA and KS reconstructions. 
For a fixed smoothing scale and weighting scheme, the noisy one-point PDF is well approximated as the convolution of the underlying (noise-free) PDF with a zero average but non-negligible variance \citep[e.g.][]{clerkin_testing_2017,barthelemy_making_2024}. 
For HSC~Y1, the noise level is high that the shape noise would dominate.
We therefore restrict the PDF test to noise-free mocks; a full treatment including noise deconvolution will be presented in future work.

\section{Discussion and conclusion} \label{sec:discussion}
We report the first application of AKRA to HSC~Y1, yielding weak-lensing convergence maps for all six survey regions.
The reconstruction pipeline was validated using realistic mock shear catalogs that reproduce the survey boundaries, masks, and spatially varying noise properties of HSC~Y1. 
AKRA performs the shear-to-convergence inversion directly, accurately recovering the mass maps, ensuring unbiased recovery of two-point and non-Gaussian statistics under realistic conditions. 
We publicly release a complete set of $\kappa$-map products for each region, including reconstructed convergence maps, survey masks, effective number-density maps ($n_{\rm eff}$), SNR maps, and power spectra. 
These products are ready for a range of scientific applications, such as cross-correlations with other tracers and non-Gaussian statistical analyses.

This work validates AKRA on real survey data and delivers accurate convergence maps for scientific analyses. The limited S/N of HSC~Y1 
also poses challenges in deriving competitive cosmological constraints.
Tomography divides sources into redshift bins to reconstruct redshift-dependent $\kappa$ maps, which improves constraints but reduces the effective number density per bin and increases shape noise. Robust cosmological analysis further requires a blinded, end-to-end likelihood that properly accounts for observational systematics (photometric redshifts, shear calibration, PSF leakage, baryonic effects, and intrinsic alignments).
In future work, we will apply AKRA to HSC~Y3 and DES to perform tomographic reconstructions and obtain cosmological constraints.

Our results show that higher-order statistics, such as skewness and the one-point PDF, are sensitive to the reconstruction method. 
AKRA preserves these non-Gaussian features under realistic conditions and provides a robust framework for field-level mass mapping. 
Non-Gaussian statistics derived from AKRA $\kappa$ maps offer a promising way to extract cosmological information beyond two-point statistics and to explore the origin of the $S_8$ tension between weak lensing and CMB results.
The reconstructed $\kappa$ field itself becomes an essential dataset that enables non-Gaussian analyses, cross-correlations, and future cosmological inference. 
We anticipate AKRA to be broadly applicable to current and upcoming weak-lensing surveys, and to support the construction of a comprehensive convergence-map data library that will provide accurate $\kappa$ maps for cosmological and cross-correlation studies.

\acknowledgments
We sincerely thank the anonymous referee for the constructive comments and suggestions.
This work was supported by the National Key R\&D Program
of China (2023YFA1607800, 2023YFA1607801, 2020YFC2201602, 2022YFF0503403), the China Manned Space Project (\#CMS-CSST-2021-A02),  China Manned Space Program (\#CMS-CSST-2025-A03), and the Fundamental Research Funds for the Central Universities.
YS acknowledges the support from NSFC Grant No. 12503004.
JY acknowledges the support from NSFC grant No. 12203084 and 
No. 12573006.
This work made use of the Gravity Supercomputer at the Department of Astronomy, Shanghai Jiao Tong University.
The results in this paper have been derived using the following packages: Numpy \cite{numpy}, Scipy \cite{SciPy},
\texttt{HEALPix} \cite{healpix}, IPython \cite{ipython} and CCL \cite{pyccl}. 

\appendix

\section{Data Release and Code Availability}
The reconstructed convergence maps, mock data products, and the AKRA code will be made publicly available at \url{https://github.com/shiyuan-1/akra_series/tree/main/AKRA_HSC}. The repository will include documentation for running AKRA, B-mode reconstruction for systematic tests, and scripts for reproducing the results presented in this paper. For further details in the AKRA series, including curved-sky implementations and other tests, is available at \url{https://github.com/shiyuan-1/akra_series/}.

\bibliographystyle{JHEP}
\bibliography{biblio.bib}

\providecommand{\href}[2]{#2}\begingroup\raggedright\begin{thebibliography}{10}

\bibitem{Bartelmann2001}
M.~Bartelmann and P.~Schneider, \emph{Weak gravitational lensing}, \href{https://doi.org/10.1016/S0370-1573(00)00082-X}{\emph{Physics Report} {\bfseries 340} (2001) 291}.

\bibitem{Refregier2003}
A.~Refregier, \emph{Weak gravitational lensing by large-scale structure},  in \emph{Annual Review of Astronomy and Astrophysics}, vol.~41, pp.~645--668, 2003, \href{https://doi.org/10.1146/annurev.astro.41.111302.102207}{DOI}.

\bibitem{Munshi2008}
D.~Munshi, P.~Valageas, L.~van Waerbeke and A.~Heavens, \emph{Cosmology with weak lensing surveys}, \href{https://doi.org/10.1016/j.physrep.2008.02.003}{\emph{Physics Reports} {\bfseries 462} (2008) 67}.

\bibitem{Fu2014}
L.P.~Fu and Z.H.~Fan, \emph{Probing the dark side of the universe with weak gravitational lensing effects}, \href{https://doi.org/10.1088/1674-4527/14/9/002}{\emph{Research in Astronomy and Astrophysics} {\bfseries 14} (2014) 1061}.

\bibitem{Kilbinger2015}
M.~Kilbinger, \emph{Cosmology with cosmic shear observations: A review}, \href{https://doi.org/10.1088/0034-4885/78/8/086901}{\emph{Reports on Progress in Physics} {\bfseries 78} (2015) }.

\bibitem{Mandelbaum2018}
R.~Mandelbaum, \emph{Weak lensing for precision cosmology}, \href{https://doi.org/10.1146/annurev-astro-081817-051928}{\emph{Annual Review of Astronomy and Astrophysics} {\bfseries 56} (2018) 393}.

\bibitem{Abbott2016}
T.~Abbott, F.B.~Abdalla, J.~Aleksić, S.~Allam, A.~Amara, D.~Bacon et~al., \emph{The dark energy survey: More than dark energy - an overview}, \href{https://doi.org/10.1093/mnras/stw641}{\emph{Monthly Notices of the Royal Astronomical Society} {\bfseries 460} (2016) 1270}.

\bibitem{Amon2022_DES_Y3}
A.~Amon, D.~Gruen, M.A.~Troxel, N.~Maccrann, S.~Dodelson, A.~Choi et~al., \emph{Dark energy survey year 3 results: Cosmology from cosmic shear and robustness to data calibration}, \href{https://doi.org/10.1103/PhysRevD.105.023514}{\emph{Physical Review D} {\bfseries 105} (2022) }.

\bibitem{Secco2022_DES_Y3}
L.F.~Secco, S.~Samuroff, E.~Krause, B.~Jain, J.~Blazek, M.~Raveri et~al., \emph{Dark energy survey year 3 results: Cosmology from cosmic shear and robustness to modeling uncertainty}, \href{https://doi.org/10.1103/PhysRevD.105.023515}{\emph{Physical Review D} {\bfseries 105} (2022) }.

\bibitem{Aihara2018}
H.~Aihara, R.~Armstrong, S.~Bickerton, J.~Bosch, J.~Coupon, H.~Furusawa et~al., \emph{First data release of the hyper suprime-cam subaru strategic program}, \href{https://doi.org/10.1093/pasj/psx081}{\emph{Publications of the Astronomical Society of Japan} {\bfseries 70} (2018) }.

\bibitem{Hikage2019_HSCY1}
C.~Hikage, M.~Oguri, T.~Hamana, S.~More, R.~Mandelbaum, M.~Takada et~al., \emph{Cosmology from cosmic shear power spectra with subaru hyper suprime-cam first-year data}, \href{https://doi.org/10.1093/pasj/psz010}{\emph{Publications of the Astronomical Society of Japan} {\bfseries 71} (2019) }.

\bibitem{Hamana2020_HSCY1}
T.~Hamana, M.~Shirasaki, S.~Miyazaki, C.~Hikage, M.~Oguri, S.~More et~al., \emph{Cosmological constraints from cosmic shear two-point correlation functions with hsc survey first-year data}, \href{https://doi.org/10.1093/pasj/psz138}{\emph{Publications of the Astronomical Society of Japan} {\bfseries 72} (2020) }.

\bibitem{Kuijken2015}
K.~Kuijken, C.~Heymans, H.~Hildebrandt, R.~Nakajima, T.~Erben, J.T.A.~De~Jong et~al., \emph{Gravitational lensing analysis of the kilo-degree survey}, \href{https://doi.org/10.1093/mnras/stv2140}{\emph{Monthly Notices of the Royal Astronomical Society} {\bfseries 454} (2015) 3500}.

\bibitem{Asgari2021_KiDS1000}
M.~Asgari, C.A.~Lin, B.~Joachimi, B.~Giblin, C.~Heymans, H.~Hildebrandt et~al., \emph{Kids-1000 cosmology: Cosmic shear constraints and comparison between two point statistics}, \href{https://doi.org/10.1051/0004-6361/202039070}{\emph{Astronomy and Astrophysics} {\bfseries 645} (2021) }.

\bibitem{Yao2024CSST}
J.~Yao, H.~Shan, R.~Li, Y.~Xu, D.~Fan, D.~Liu et~al., \emph{Csst wl preparation i: forecast the impact from non-gaussian covariances and requirements on systematics control}, \href{https://doi.org/10.1093/mnras/stad3563}{\emph{Monthly Notices of the Royal Astronomical Society} {\bfseries 527} (2024) 5206}.

\bibitem{Laureijs2011}
R.~Laureijs, J.~Amiaux, S.~Arduini, J.L.~AuguÅ res, J.~Brinchmann, R.~Cole et~al., \emph{Euclid definition study report}, \href{https://doi.org/arXiv:1110.3193}{\emph{Arxiv} (2011) }.

\bibitem{Euclid2022}
C.~Euclid, Y.~Mellier, Abdurro'uf, J.A.A.~Barroso, A.~Achucarro, J.~Adamek et~al., \emph{Euclid. i. overview of the euclid mission}, .

\bibitem{LSST2009}
L.S.~Collaboration, P.A.~Abell, J.~Allison, S.F.~Anderson, J.R.~Andrew, J.R.P.~Angel et~al., \emph{Lsst science book, version 2.0},  December 01, 2009, 2009.
\newblock 10.48550/arXiv.0912.0201.

\bibitem{Ivezi2020}
Å.~Ivezić, S.M.~Kahn, J.A.~Tyson, B.~Abel, E.~Acosta, R.~Allsman et~al., \emph{Lsst: From science drivers to reference design and anticipated data products}, \href{https://doi.org/10.3847/1538-4357/ab042c}{\emph{Astrophysical Journal} {\bfseries 873} (2019) }.

\bibitem{Spergel2015}
D.~Spergel, N.~Gehrels, C.~Baltay, D.~Bennett, J.~Breckinridge, M.~Donahue et~al., \emph{Wide-field infrarred survey telescope-astrophysics focused telescope assets wfirst-afta 2015 report},  March 01, 2015, 2015.
\newblock 10.48550/arXiv.1503.03757.

\bibitem{Gong2019}
Y.~Gong, X.~Liu, Y.~Cao, X.~Chen, Z.~Fan, R.~Li et~al., \emph{Cosmology from the chinese space station optical survey (css-os)}, \href{https://doi.org/10.3847/1538-4357/ab391e}{\emph{ASTROPHYSICAL JOURNAL} {\bfseries 883} (2019) }.

\bibitem{Zhan2021}
H.~Zhan, \emph{The wide-field multiband imaging and slitless spectroscopy survey to be carried out by the survey space telescope of china manned space program}, \href{https://doi.org/10.1360/TB-2021-0016}{\emph{Kexue Tongbao/Chinese Science Bulletin} {\bfseries 66} (2021) 1290}.

\bibitem{Shan2018}
H.~{Shan}, X.~{Liu}, H.~{Hildebrandt}, C.~{Pan}, N.~{Martinet}, Z.~{Fan} et~al., \emph{{KiDS-450: cosmological constraints from weak lensing peak statistics - I. Inference from analytical prediction of high signal-to-noise ratio convergence peaks}}, \href{https://doi.org/10.1093/mnras/stx2837}{\emph{Monthly Notices of the Royal Astronomical Society} {\bfseries 474} (2018) 1116} [\href{https://arxiv.org/abs/1709.07651}{{\ttfamily 1709.07651}}].

\bibitem{Martinet2018}
N.~{Martinet}, P.~{Schneider}, H.~{Hildebrandt}, H.~{Shan}, M.~{Asgari}, J.P.~{Dietrich} et~al., \emph{{KiDS-450: cosmological constraints from weak-lensing peak statistics - II: Inference from shear peaks using N-body simulations}}, \href{https://doi.org/10.1093/mnras/stx2793}{\emph{Monthly Notices of the Royal Astronomical Society} {\bfseries 474} (2018) 712} [\href{https://arxiv.org/abs/1709.07678}{{\ttfamily 1709.07678}}].

\bibitem{Liu2023}
D.Z.~{Liu}, X.M.~{Meng}, X.Z.~{Er}, Z.H.~{Fan}, M.~{Kilbinger}, G.L.~{Li} et~al., \emph{{Potential scientific synergies in weak lensing studies between the CSST and Euclid space probes}}, \href{https://doi.org/10.1051/0004-6361/202243978}{\emph{Astronomy \& Astrophysics} {\bfseries 669} (2023) A128} [\href{https://arxiv.org/abs/2210.16341}{{\ttfamily 2210.16341}}].

\bibitem{VanWaerbeke2013}
L.~{Van Waerbeke}, J.~{Benjamin}, T.~{Erben}, C.~{Heymans}, H.~{Hildebrandt}, H.~{Hoekstra} et~al., \emph{{CFHTLenS: mapping the large-scale structure with gravitational lensing}}, \href{https://doi.org/10.1093/mnras/stt971}{\emph{Monthly Notices of the Royal Astronomical Society} {\bfseries 433} (2013) 3373} [\href{https://arxiv.org/abs/1303.1806}{{\ttfamily 1303.1806}}].

\bibitem{Petri2015}
A.~{Petri}, J.~{Liu}, Z.~{Haiman}, M.~{May}, L.~{Hui} and J.M.~{Kratochvil}, \emph{{Emulating the CFHTLenS weak lensing data: Cosmological constraints from moments and Minkowski functionals}}, \href{https://doi.org/10.1103/PhysRevD.91.103511}{\emph{Physical Review D} {\bfseries 91} (2015) 103511} [\href{https://arxiv.org/abs/1503.06214}{{\ttfamily 1503.06214}}].

\bibitem{Chang2018}
C.~{Chang}, A.~{Pujol}, B.~{Mawdsley}, D.~{Bacon}, J.~{Elvin-Poole}, P.~{Melchior} et~al., \emph{{Dark Energy Survey Year 1 results: curved-sky weak lensing mass map}}, \href{https://doi.org/10.1093/mnras/stx3363}{\emph{Monthly Notices of the Royal Astronomical Society} {\bfseries 475} (2018) 3165} [\href{https://arxiv.org/abs/1708.01535}{{\ttfamily 1708.01535}}].

\bibitem{Kratochvil2012}
J.M.~{Kratochvil}, E.A.~{Lim}, S.~{Wang}, Z.~{Haiman}, M.~{May} and K.~{Huffenberger}, \emph{{Probing cosmology with weak lensing Minkowski functionals}}, \href{https://doi.org/10.1103/PhysRevD.85.103513}{\emph{Physical Review D} {\bfseries 85} (2012) 103513} [\href{https://arxiv.org/abs/1109.6334}{{\ttfamily 1109.6334}}].

\bibitem{Vicinanza2019}
M.~{Vicinanza}, V.F.~{Cardone}, R.~{Maoli}, R.~{Scaramella}, X.~{Er} and I.~{Tereno}, \emph{{Minkowski functionals of convergence maps and the lensing figure of merit}}, \href{https://doi.org/10.1103/PhysRevD.99.043534}{\emph{Physical Review D} {\bfseries 99} (2019) 043534} [\href{https://arxiv.org/abs/1905.00410}{{\ttfamily 1905.00410}}].

\bibitem{Zurcher2021}
D.~Zürcher, J.~Fluri, R.~Sgier, T.~Kacprzak and A.~Refregier, \emph{Cosmological forecast for non-gaussian statistics in large-scale weak lensing surveys}, \href{https://doi.org/10.1088/1475-7516/2021/01/028}{\emph{Journal of Cosmology and Astroparticle Physics} {\bfseries 2021} (2021) 028–028}.

\bibitem{Cheng2020}
S.~Cheng, Y.S.~Ting, B.~Menard and J.~Bruna, \emph{A new approach to observational cosmology using the scattering transform}, \href{https://doi.org/10.1093/MNRAS/STAA3165}{\emph{MONTHLY NOTICES OF THE ROYAL ASTRONOMICAL SOCIETY} {\bfseries 499} (2020) 5902}.

\bibitem{Cheng2021}
S.~Cheng and B.~Ménard, \emph{Weak lensing scattering transform: dark energy and neutrino mass sensitivity}, \href{https://doi.org/10.1093/mnras/stab2102}{\emph{MONTHLY NOTICES OF THE ROYAL ASTRONOMICAL SOCIETY} {\bfseries 507} (2021) 1012}.

\bibitem{Takada2003}
M.~Takada and B.~Jain, \emph{Three-point correlations in weak lensing surveys: Model predictions and applications}, \href{https://doi.org/10.1046/j.1365-8711.2003.06868.x}{\emph{MONTHLY NOTICES OF THE ROYAL ASTRONOMICAL SOCIETY} {\bfseries 344} (2003) 857}.

\bibitem{Gupta2018}
A.~{Gupta}, J.M.Z.~{Matilla}, D.~{Hsu} and Z.~{Haiman}, \emph{{Non-Gaussian information from weak lensing data via deep learning}}, \href{https://doi.org/10.1103/PhysRevD.97.103515}{\emph{Physical Review D} {\bfseries 97} (2018) 103515} [\href{https://arxiv.org/abs/1802.01212}{{\ttfamily 1802.01212}}].

\bibitem{Ribli2019a}
D.~{Ribli}, B.{\'A}.~{Pataki} and I.~{Csabai}, \emph{{An improved cosmological parameter inference scheme motivated by deep learning}}, \href{https://doi.org/10.1038/s41550-018-0596-8}{\emph{Nature Astronomy} {\bfseries 3} (2019) 93} [\href{https://arxiv.org/abs/1806.05995}{{\ttfamily 1806.05995}}].

\bibitem{Fluri2022}
J.~{Fluri}, T.~{Kacprzak}, A.~{Lucchi}, A.~{Schneider}, A.~{Refregier} and T.~{Hofmann}, \emph{{Full w CDM analysis of KiDS-1000 weak lensing maps using deep learning}}, \href{https://doi.org/10.1103/PhysRevD.105.083518}{\emph{Physical Review D} {\bfseries 105} (2022) 083518} [\href{https://arxiv.org/abs/2201.07771}{{\ttfamily 2201.07771}}].

\bibitem{LuTianhuan2023}
T.~{Lu}, Z.~{Haiman} and X.~{Li}, \emph{{Cosmological constraints from HSC survey first-year data using deep learning}}, \href{https://doi.org/10.1093/mnras/stad686}{\emph{Monthly Notices of the Royal Astronomical Society} {\bfseries 521} (2023) 2050} [\href{https://arxiv.org/abs/2301.01354}{{\ttfamily 2301.01354}}].

\bibitem{Zhou2024}
A.J.~Zhou, X.~Li, S.~Dodelson and R.~Mandelbaum, \emph{Accurate field-level weak lensing inference for precision cosmology}, \href{https://doi.org/10.1103/PhysRevD.110.023539}{\emph{Phys. Rev. D} {\bfseries 110} (2024) 023539}.

\bibitem{Zeghal2025}
J.~Zeghal, D.~Lanzieri, F.~Lanusse, A.~Boucaud, G.~Louppe, E.~Aubourg et~al., \emph{Simulation-based inference benchmark for weak lensing cosmology}, \href{https://doi.org/10.1051/0004-6361/202452410}{\emph{Astronomy and Astrophysics} {\bfseries 699} (2025) A327}.

\bibitem{Baldauf2010}
T.~{Baldauf}, R.E.~{Smith}, U.~{Seljak} and R.~{Mandelbaum}, \emph{{Algorithm for the direct reconstruction of the dark matter correlation function from weak lensing and galaxy clustering}}, \href{https://doi.org/10.1103/PhysRevD.81.063531}{\emph{PHYSICAL REVIEW D} {\bfseries 81} (2010) 063531} [\href{https://arxiv.org/abs/0911.4973}{{\ttfamily 0911.4973}}].

\bibitem{MacCrann2020}
N.~{MacCrann}, J.~{Blazek}, B.~{Jain} and E.~{Krause}, \emph{{Controlling and leveraging small-scale information in tomographic galaxy-galaxy lensing}}, \href{https://doi.org/10.1093/mnras/stz2761}{\emph{Mon. Not. Roy. Astron. Soc} {\bfseries 491} (2020) 5498} [\href{https://arxiv.org/abs/1903.07101}{{\ttfamily 1903.07101}}].

\bibitem{Park2021}
Y.~{Park}, E.~{Rozo} and E.~{Krause}, \emph{{Localizing Transformations of the Galaxy-Galaxy Lensing Observable}}, \href{https://doi.org/10.1103/PhysRevLett.126.021301}{\emph{Physical Review Letters} {\bfseries 126} (2021) 021301} [\href{https://arxiv.org/abs/2004.07504}{{\ttfamily 2004.07504}}].

\bibitem{Prat2022}
J.~{Prat}, J.~{Blazek}, C.~{S{\'a}nchez}, I.~{Tutusaus}, S.~{Pandey}, J.~{Elvin-Poole} et~al., \emph{{Dark energy survey year 3 results: High-precision measurement and modeling of galaxy-galaxy lensing}}, \href{https://doi.org/10.1103/PhysRevD.105.083528}{\emph{PHYSICAL REVIEW D} {\bfseries 105} (2022) 083528} [\href{https://arxiv.org/abs/2105.13541}{{\ttfamily 2105.13541}}].

\bibitem{Kaiser1993}
N.~Kaiser and G.~Squires, \emph{Mapping the dark matter with weak gravitational lensing}, \href{https://doi.org/10.1086/172297}{\emph{Astrophysical Journal} {\bfseries 404} (1993) 441}.

\bibitem{Vikram2015}
V.~Vikram, C.~Chang, B.~Jain, D.~Bacon, A.~Amara, M.R.~Becker et~al., \emph{Wide-field lensing mass maps from dark energy survey science verification data: Methodology and detailed analysis}, \href{https://doi.org/10.1103/PhysRevD.92.022006}{\emph{Physical Review D - Particles, Fields, Gravitation and Cosmology} {\bfseries 92} (2015) }.

\bibitem{Gatti2022}
M.~Gatti, B.~Jain, C.~Chang, M.~Raveri, D.~Zürcher, L.~Secco et~al., \emph{Dark energy survey year 3 results: Cosmology with moments of weak lensing mass maps}, \href{https://doi.org/10.1103/PhysRevD.106.083509}{\emph{PHYSICAL REVIEW D} {\bfseries 106} (2022) }.

\bibitem{Alsing2016}
J.~{Alsing}, A.~{Heavens}, A.H.~{Jaffe}, A.~{Kiessling}, B.~{Wandelt} and T.~{Hoffmann}, \emph{{Hierarchical cosmic shear power spectrum inference}}, \href{https://doi.org/10.1093/mnras/stv2501}{\emph{Monthly Notices of the Royal Astronomical Society} {\bfseries 455} (2016) 4452} [\href{https://arxiv.org/abs/1505.07840}{{\ttfamily 1505.07840}}].

\bibitem{Alsing2017}
J.~{Alsing}, A.~{Heavens} and A.H.~{Jaffe}, \emph{{Cosmological parameters, shear maps and power spectra from CFHTLenS using Bayesian hierarchical inference}}, \href{https://doi.org/10.1093/mnras/stw3161}{\emph{Monthly Notices of the Royal Astronomical Society} {\bfseries 466} (2017) 3272} [\href{https://arxiv.org/abs/1607.00008}{{\ttfamily 1607.00008}}].

\bibitem{Porqueres2022}
N.~{Porqueres}, A.~{Heavens}, D.~{Mortlock} and G.~{Lavaux}, \emph{{Lifting weak lensing degeneracies with a field-based likelihood}}, \href{https://doi.org/10.1093/mnras/stab3234}{\emph{Monthly Notices of the Royal Astronomical Society} {\bfseries 509} (2022) 3194} [\href{https://arxiv.org/abs/2108.04825}{{\ttfamily 2108.04825}}].

\bibitem{Jeffrey2018}
N.~{Jeffrey}, F.B.~{Abdalla}, O.~{Lahav}, F.~{Lanusse}, J.L.~{Starck}, A.~{Leonard} et~al., \emph{{Improving weak lensing mass map reconstructions using Gaussian and sparsity priors: application to DES SV}}, \href{https://doi.org/10.1093/mnras/sty1252}{\emph{Monthly Notices of the Royal Astronomical Society} {\bfseries 479} (2018) 2871} [\href{https://arxiv.org/abs/1801.08945}{{\ttfamily 1801.08945}}].

\bibitem{Leonard2014}
A.~{Leonard}, F.~{Lanusse} and J.-L.~{Starck}, \emph{{GLIMPSE: accurate 3D weak lensing reconstructions using sparsity}}, \href{https://doi.org/10.1093/mnras/stu273}{\emph{Monthly Notices of the Royal Astronomical Society} {\bfseries 440} (2014) 1281} [\href{https://arxiv.org/abs/1308.1353}{{\ttfamily 1308.1353}}].

\bibitem{Price2019}
M.A.~{Price}, J.D.~{McEwen}, X.~{Cai}, T.D.~{Kitching} and {LSST Dark Energy Science Collaboration}, \emph{{Sparse Bayesian mass mapping with uncertainties: peak statistics and feature locations}}, \href{https://doi.org/10.1093/mnras/stz2373}{\emph{Monthly Notices of the Royal Astronomical Society} {\bfseries 489} (2019) 3236} [\href{https://arxiv.org/abs/1812.04018}{{\ttfamily 1812.04018}}].

\bibitem{Fiedorowicz2022b}
P.~{Fiedorowicz}, E.~{Rozo} and S.S.~{Boruah}, \emph{{KaRMMa 2.0 -- Kappa Reconstruction for Mass Mapping}}, \href{https://doi.org/10.48550/arXiv.2210.12280}{\emph{arXiv e-prints} (2022) arXiv:2210.12280} [\href{https://arxiv.org/abs/2210.12280}{{\ttfamily 2210.12280}}].

\bibitem{Fiedorowicz2022a}
P.~{Fiedorowicz}, E.~{Rozo}, S.S.~{Boruah}, C.~{Chang} and M.~{Gatti}, \emph{{KaRMMa - kappa reconstruction for mass mapping}}, \href{https://doi.org/10.1093/mnras/stac468}{\emph{Monthly Notices of the Royal Astronomical Society} {\bfseries 512} (2022) 73} [\href{https://arxiv.org/abs/2105.14699}{{\ttfamily 2105.14699}}].

\bibitem{Starck2006}
J.L.~{Starck}, S.~{Pires} and A.~{R{\'e}fr{\'e}gier}, \emph{{Weak lensing mass reconstruction using wavelets}}, \href{https://doi.org/10.1051/0004-6361:20052997}{\emph{Astronomy \& Astrophysics} {\bfseries 451} (2006) 1139} [\href{https://arxiv.org/abs/astro-ph/0503373}{{\ttfamily astro-ph/0503373}}].

\bibitem{Starck2021}
J.L.~{Starck}, K.E.~{Themelis}, N.~{Jeffrey}, A.~{Peel} and F.~{Lanusse}, \emph{{Weak-lensing mass reconstruction using sparsity and a Gaussian random field}}, \href{https://doi.org/10.1051/0004-6361/202039451}{\emph{Astronomy \& Astrophysics} {\bfseries 649} (2021) A99} [\href{https://arxiv.org/abs/2102.04127}{{\ttfamily 2102.04127}}].

\bibitem{Shi2024}
Y.~Shi, P.~Zhang, Z.~Sun and Y.~Wang, \emph{Accurate kappa reconstruction algorithm for masked shear catalog}, \href{https://doi.org/10.1103/PhysRevD.109.123530}{\emph{PHYSICAL REVIEW D} {\bfseries 109} (2024) 123530}.

\bibitem{Shi2025}
Y.~Shi, P.~Zhang, F.~Deng, S.~Zhou, H.~Cai, J.~Yao et~al., \emph{Akra 2.0: Accurate kappa reconstruction algorithm for masked shear catalog}, \href{https://doi.org/10.1088/1475-7516/2025/07/038}{\emph{JOURNAL OF COSMOLOGY AND ASTROPARTICLE PHYSICS} {\bfseries 2025} (2025) 038}.

\bibitem{Mandelbaum2018_HSCY1_data}
R.~Mandelbaum, H.~Miyatake, T.~Hamana, M.~Oguri, M.~Simet, R.~Armstrong et~al., \emph{The first-year shear catalog of the subaru hyper suprime-cam subaru strategic program survey}, \href{https://doi.org/10.1093/pasj/psx130}{\emph{Publications of the Astronomical Society of Japan} {\bfseries 70} (2018) }.

\bibitem{Liu2023_HSC}
X.~Liu, S.~Yuan, C.~Pan, T.~Zhang, Q.~Wang and Z.~Fan, \emph{Cosmological studies from hsc-ssp tomographic weak-lensing peak abundances}, \href{https://doi.org/10.1093/mnras/stac2971}{\emph{Monthly Notices of the Royal Astronomical Society} {\bfseries 519} (2023) 594}.

\bibitem{Lu2023_HSC}
T.~Lu, Z.~Haiman and X.~Li, \emph{Cosmological constraints from hsc survey first-year data using deep learning}, \href{https://doi.org/10.1093/mnras/stad686}{\emph{Monthly Notices of the Royal Astronomical Society} {\bfseries 521} (2023) 2050}.

\bibitem{Thiele2023_HSCY1_PDF}
L.~Thiele, G.A.~Marques, J.~Liu and M.~Shirasaki, \emph{Cosmological constraints from the subaru hyper suprime-cam year 1 shear catalogue lensing convergence probability distribution function}, \href{https://doi.org/10.1103/PhysRevD.108.123526}{\emph{Physical Review D} {\bfseries 108} (2023) }.

\bibitem{Grandón2024_HSC}
D.~Grandón, G.A.~Marques, L.~Thiele, S.~Cheng, M.~Shirasaki and J.~Liu, \emph{Impact of baryonic feedback on hsc-y1 weak lensing non-gaussian statistics}, \href{https://doi.org/10.1103/PhysRevD.110.103539}{\emph{Physical Review D} {\bfseries 110} (2024) }.

\bibitem{Marques2024_HSC}
G.A.~Marques, J.~Liu, M.~Shirasaki, L.~Thiele, D.~Grandón, K.M.~Huffenberger et~al., \emph{Cosmology from weak lensing peaks and minima with subaru hyper suprime-cam survey first-year data}, \href{https://doi.org/10.1093/mnras/stae098}{\emph{Monthly Notices of the Royal Astronomical Society} {\bfseries 528} (2024) 4513}.

\bibitem{Novaes2024_HSC}
C.P.~Novaes, L.~Thiele, J.~Armijo, S.~Cheng, J.A.~Cowell, G.A.~Marques et~al., \emph{Cosmology from hsc y1 weak lensing with combined higher-order statistics and simulation-based inference}, {\emph{Cosmology from HSC Y1 Weak Lensing with Combined Higher-Order Statistics and Simulation-based Inference} (2024) }.

\bibitem{Armijo2025_HSC}
J.~Armijo, G.A.~Marques, C.P.~Novaes, L.~Thiele, J.A.~Cowell, D.~Grandón et~al., \emph{Cosmological constraints using minkowski functionals from the first year data of the hyper suprime-cam}, \href{https://doi.org/10.1093/mnras/staf257}{\emph{Monthly Notices of the Royal Astronomical Society} {\bfseries 537} (2025) 3553}.

\bibitem{Cheng2025_HSC}
S.~Cheng, G.A.~Marques, D.~Grandón, L.~Thiele, M.~Shirasaki, B.~Ménard et~al., \emph{Cosmological constraints from weak lensing scattering transform using hsc y1 data}, \href{https://doi.org/10.1088/1475-7516/2025/01/006}{\emph{Journal of Cosmology and Astroparticle Physics} {\bfseries 2025} (2025) }.

\bibitem{Tanaka2018}
M.~Tanaka, J.~Coupon, B.C.~Hsieh, S.~Mineo, A.J.~Nishizawa, J.~Speagle et~al., \emph{Photometric redshifts for hyper suprime-cam subaru strategic program data release 1}, \href{https://doi.org/10.1093/pasj/psx077}{\emph{Publications of the Astronomical Society of Japan} {\bfseries 70} (2018) }.

\bibitem{hikage_cosmology_2019}
C.~Hikage, M.~Oguri, T.~Hamana, S.~More, R.~Mandelbaum, M.~Takada et~al., \emph{Cosmology from cosmic shear power spectra with {Subaru} {Hyper} {Suprime}-{Cam} first-year data}, \href{https://doi.org/10.1093/pasj/psz010}{\emph{Publications of the Astronomical Society of Japan} {\bfseries 71} (2019) 43}.

\bibitem{hamana_cosmological_2020}
T.~Hamana, M.~Shirasaki, S.~Miyazaki, C.~Hikage, M.~Oguri, S.~More et~al., \emph{Cosmological constraints from cosmic shear two-point correlation functions with {HSC} survey first-year data}, \href{https://doi.org/10.1093/pasj/psz138}{\emph{Publications of the Astronomical Society of Japan} {\bfseries 72} (2020) }.

\bibitem{2025SCPMA..6889512C}
Z.~{Chen}, Y.~{Yu}, J.~{Han} and Y.~{Jing}, \emph{{CSST cosmological emulator I: Matter power spectrum emulation with one percent accuracy to k = 10h Mpc$^{‑1}$}}, \href{https://doi.org/10.1007/s11433-025-2671-0}{\emph{Science China Physics, Mechanics, and Astronomy} {\bfseries 68} (2025) 289512} [\href{https://arxiv.org/abs/2502.11160}{{\ttfamily 2502.11160}}].

\bibitem{2025SCPMA..6809511H}
J.~{Han}, M.~{Li}, W.~{Jiang}, Z.~{Chen}, H.~{Wang}, C.~{Wei} et~al., \emph{{The Jiutian simulations for the CSST extra-galactic surveys}}, \href{https://doi.org/10.1007/s11433-025-2712-1}{\emph{Science China Physics, Mechanics, and Astronomy} {\bfseries 68} (2025) 109511} [\href{https://arxiv.org/abs/2503.21368}{{\ttfamily 2503.21368}}].

\bibitem{2025SCPMA..6809513C}
Z.~{Chen} and Y.~{Yu}, \emph{{CSST cosmological emulator II: Generalized accurate halo mass function emulation}}, \href{https://doi.org/10.1007/s11433-025-2764-x}{\emph{Science China Physics, Mechanics, and Astronomy} {\bfseries 68} (2025) 109513} [\href{https://arxiv.org/abs/2506.09688}{{\ttfamily 2506.09688}}].

\bibitem{2025arXiv250604671Z}
S.~{Zhou}, Z.~{Chen} and Y.~{Yu}, \emph{{CSST Cosmological Emulator III: Hybrid Lagrangian Bias Expansion Emulation of Galaxy Clustering}}, \href{https://doi.org/10.48550/arXiv.2506.04671}{\emph{arXiv e-prints} (2025) arXiv:2506.04671} [\href{https://arxiv.org/abs/2506.04671}{{\ttfamily 2506.04671}}].

\bibitem{Shirasaki2019}
M.~{Shirasaki}, T.~{Hamana}, M.~{Takada}, R.~{Takahashi} and H.~{Miyatake}, \emph{{Mock galaxy shape catalogues in the Subaru Hyper Suprime-Cam Survey}}, \href{https://doi.org/10.1093/mnras/stz791}{\emph{MONTHLY NOTICES OF THE ROYAL ASTRONOMICAL SOCIETY} {\bfseries 486} (2019) 52} [\href{https://arxiv.org/abs/1901.09488}{{\ttfamily 1901.09488}}].

\bibitem{Bernardeau1997}
F.~Bernardeau, L.~van Waerbeke and Y.~Mellier, \emph{Weak lensing statistics as a probe of {OMEGA} and power spectrum}, \href{https://doi.org/10.48550/arXiv.astro-ph/9609122}{\emph{Astronomy and Astrophysics} {\bfseries 322} (1997) 1}.

\bibitem{barthelemy_probability_2021}
A.~Barthelemy, S.~Codis and F.~Bernardeau, \emph{Probability distribution function of the aperture mass field with large deviation theory}, \href{https://doi.org/10.1093/mnras/stab818}{\emph{Monthly Notices of the Royal Astronomical Society} {\bfseries 503} (2021) 5204}.

\bibitem{boyle_nuw_2021}
A.~Boyle, C.~Uhlemann, O.~Friedrich, A.~Barthelemy, S.~Codis, F.~Bernardeau et~al., \emph{Nuw {CDM} cosmology from the weak lensing convergence {PDF}}, \href{https://doi.org/10.1093/mnras/stab1381}{\emph{Monthly Notices of the Royal Astronomical Society} {\bfseries 505} (2021) 2886}.

\bibitem{barthelemy_making_2024}
A.~Barthelemy, A.~Halder, Z.~Gong and C.~Uhlemann, \emph{Making the leap. {Part} {I}. {Modelling} the reconstructed lensing convergence {PDF} from cosmic shear with survey masks and systematics}, \href{https://doi.org/10.1088/1475-7516/2024/03/060}{\emph{Journal of Cosmology and Astroparticle Physics} {\bfseries 2024} (2024) 060}.

\bibitem{taruya_lognormal_2002}
A.~Taruya, M.~Takada, T.~Hamana, I.~Kayo and T.~Futamase, \emph{Lognormal {Property} of {Weak}‐{Lensing} {Fields}}, \href{https://doi.org/10.1086/340048}{\emph{The Astrophysical Journal} {\bfseries 571} (2002) 638}.

\bibitem{clerkin_testing_2017}
L.~Clerkin, D.~Kirk, M.~Manera, O.~Lahav, F.~Abdalla, A.~Amara et~al., \emph{Testing the lognormality of the galaxy and weak lensing convergence distributions from {Dark} {Energy} {Survey} maps}, \href{https://doi.org/10.1093/mnras/stw2106}{\emph{Monthly Notices of the Royal Astronomical Society} {\bfseries 466} (2017) 1444}.

\bibitem{numpy}
S.~van~der Walt, S.C.~Colbert and G.~Varoquaux, \emph{The numpy array: A structure for efficient numerical computation}, \href{https://doi.org/10.1109/MCSE.2011.37}{\emph{Computing in Science Engineering} {\bfseries 13} (2011) 22}.

\bibitem{SciPy}
P.~Virtanen, R.~Gommers, T.E.~Oliphant, M.~Haberland, T.~Reddy, D.~Cournapeau et~al., \emph{{{SciPy} 1.0: Fundamental Algorithms for Scientific Computing in Python}}, \href{https://doi.org/10.1038/s41592-019-0686-2}{\emph{Nature Methods} {\bfseries 17} (2020) 261}.

\bibitem{healpix}
K.M.~{G{\'o}rski}, E.~{Hivon}, A.J.~{Banday}, B.D.~{Wand elt}, F.K.~{Hansen}, M.~{Reinecke} et~al., \emph{{HEALPix: A Framework for High-Resolution Discretization and Fast Analysis of Data Distributed on the Sphere}}, \href{https://doi.org/10.1086/427976}{\emph{Astrophysical Journal} {\bfseries 622} (2005) 759} [\href{https://arxiv.org/abs/astro-ph/0409513}{{\ttfamily astro-ph/0409513}}].

\bibitem{ipython}
F.~Perez and B.E.~Granger, \emph{Ipython: A system for interactive scientific computing}, \href{https://doi.org/10.1109/MCSE.2007.53}{\emph{Computing in Science Engineering} {\bfseries 9} (2007) 21}.

\bibitem{pyccl}
N.E.~{Chisari}, D.~{Alonso}, E.~{Krause}, C.D.~{Leonard}, P.~{Bull}, J.~{Neveu} et~al., \emph{{Core Cosmology Library: Precision Cosmological Predictions for LSST}}, \href{https://doi.org/10.3847/1538-4365/ab1658}{\emph{Astrophysical Journals} {\bfseries 242} (2019) 2} [\href{https://arxiv.org/abs/1812.05995}{{\ttfamily 1812.05995}}].

\end{thebibliography}\endgroup

\end{document}